# Black Phosphorus-Polymer Composites for Pulsed Lasers

*Haoran Mu, Shenghuang Lin, Zhongchi Wang, Si Xiao, Pengfei Li, Yao Chen, Han Zhang, Haifeng Bao, Shu Ping Lau, Chunxu Pan, Dianyuan Fan, Qiaoliang Bao**

H. Mu, Dr. S. Lin, P. Li, Y. Chen, Prof. Q. Bao
Institute of Functional Nano and Soft Materials (FUNSOM), Jiangsu Key Laboratory for Carbon-Based Functional Materials and Devices, and Collaborative Innovation Center of Suzhou Nano Science and Technology, Soochow University, Suzhou 215123, P. R. China
(E-mail: qlbao@suda.edu.cn)

Dr. S. Lin, Prof. S. P. Lau
Department of Applied Physics, The Hong Kong Polytechnic University, Hung Hom, Hong Kong SAR, China

Z. Wang, Prof. C. Pan
School of Physical and Technology, Wuhan University, Wuhan 430072, P. R. China

Prof. S. Xiao
Institute of Super-Microstructure and Ultrafast Process in Advanced Materials, School of Physics and Electronics, Hunan Key Laboratory for Super-Microstructure and Ultrafast Process, Central South University, Changsha 410083, China

Prof. H. Zhang, Prof. D. Fan
SZU-NUS Collaborative Innovation Centre for Optoelectronic Science & Technology, and Key Laboratory of Optoelectronic Devices and Systems of Ministry of Education and Guangdong Province, Shenzhen University, Shenzhen, China

Prof. H. Bao
School of Materials Science and Engineering, Wuhan Textile University, Wuhan 430200, China

**\***H. Mu, S. Lin and Z. Wang contributed equally to this work.

\**Correspondence to*: Prof. Q. Bao (E-mail: qlbao@suda.edu.cn)

**ABSTRACT**

Black phosphorus is a very promising material for telecommunication due to its direct bandgap and strong resonant absorption in near-infrared wavelength range. However, ultrafast nonlinear photonic applications relying on the ultrafast photo-carrier dynamics as well as optical nonlinearity in black phosphorus remain unexplored. In this work, we investigate nonlinear optical properties of solution exfoliated BP and demonstrate the usage of BP as a new saturable absorber for high energy pulse generation in fiber laser. In order to avoid the oxidization and degradation of BP, we encapsulated BP by polymer matrix which is optically transparent in the spectrum range of interest to form a composite. Two fabrication approaches were demonstrated to produce BP-polymer composite films which were further incorporated into fiber laser cavity as nonlinear media. BP shows very fast carrier dynamics and BP-polymer composite has a modulation depth of 10.6%. A highly stable Q-switched pulse generation was achieved and the single pulse energy of ~194 nJ was demonstrated. The ease of handling of such black phosphorus-polymer composite thin films affords new opportunities for wider applications such as optical sensing, signal processing and light modulation.

# 1. Introduction

The exploration of atomically thin layered materials such as graphene and transition metal dichalcogenides (TMDs) offered many new opportunities for optoelectronic and photonic applications.[1–3] Owing to its broadband absorption, ultrafast dynamic response and large optical nonlinearity, graphene has been demonstrated to be a new nonlinear optical material for the generation and modulation of laser pulses,[4,5] which could find important applications for telecommunications and optical sensing. However, the relative weak absorption in monolayer graphene (2.3 % of incident light) gives a small absolute value (approximately 1%) of optical modulation depth,[6,7] which brings certain limitation for nonlinear photonics at specific wavelength. TMDs have a high resonant absorption up to >20%,[8,9] but the optical response mainly lies in visible range due to a moderate band gap (1 eV for bulk and 2 eV for monolayer)[10]. It is noteworthy that great technological demands for optical communications exist mainly in the infrared wavelength range around 1500 nm (0.8 eV). To this point, black phosphorus (BP),[112] a new member of the layered-material family with band gap from 0.3 eV (bulk) to 1.5 eV (monolayer),[13-15] can bridge the gap between zero-gap graphene and relatively large bandgap TMDs for infrared photonics and optoelectronics.

Because of the unique orthorhombic crystal structure[16], BP shows tunable optical properties which are very sensitive to thickness, doping and light polarization.[13,17,18] Unlike TMDs, in which multilayers have an indirect bandgap,[10] BP always has a direct bandgap for all thicknesses, leading to extraordinary light emission[19] and efficient photo-electrical conversion[20]. Another important consequence from the direct bandgap is ultrafast carrier dynamics, which is a significant benefit for ultrafast photonics and high frequency optoelectronics.[21,22] Despite of the few demonstrations for BP-based photo-detectors[18, 23-25] and photovoltaic device[26], ultrafast nonlinear photonic applications relying on the photo-carrier dynamics as well as optical nonlinearity in BP remain unexplored[26]. In this work, we investigate ultrafast optical properties of BP and demonstrate the usage of BP as a new nonlinear optical material for high energy pulse generation. We have noticed that most of the experimental works on BP were performed under the condition of vacuum or inert gas [11,12,23-25,27,28] due to the instability of BP in the atmosphere. In order to avoid the oxidization of BP, we encapsulated BP by polymer matrix which is optically transparent in the spectrum range of interest to form a composite. The ease of handling of such BP-polymer composite thin films affords new opportunities for wide photonic applications.

## 2. Results and discussion

### 2.1 BP solution characterization

A solution-based mechanical exfoliation process is used to produce large quantities of multilayer BP nanoflakes, which is described in the experimental section in detail. Figure 1a shows the representative transmission electron microscopy (TEM) image of the as-produced BP nanocrystals, which appear to be layered thin sheets with the size up to 100 nm. The high resolution TEM image in Figure 1b clearly resolves the microstructure of BP, in which the lattice space of 0.26 nm corresponds to (040) facets. The BP samples have quite good crystallinity, which is further confirmed by the selective area diffraction (SAED) pattern shown in the inset of Figure 1b. The atomic force microscopy (AFM) measurements were performed to investigate the morphology of BP nanoflakes. It is found that the thickness ranges from 4 nm to 25 nm, indicating that our sample consists of mostly multilayer BP. Figure 1d shows the Raman spectrum of BP sample, which depicts three characteristic peaks at 360.3 cm$^{-1}$, 437.1 cm$^{-1}$, and 464.8 cm$^{-1}$, corresponding to $A_g^1$, $B_g^2$ and $A_g^2$ vibration modes[29], respectively.

The exfoliated BP dispersed in ethanol appears to be brown in color (inset of Figure 1e). The UV-visible-infrared absorption spectrum of BP solution is depicted in Figure 1e. It is interesting to observe a resonant peak centered at 1500 nm, across a few telecommunication bands from 1420 nm to 1630 nm. The enhanced light absorption peak is leaded by the optical band gap of four- and five-layers thick BP flakes in the solution[30], which indicates the application potential of BP solution in near-infra wavelength range. Pump-probe experiments at 1550 nm (telecommunication C band) were carried out to study the carrier dynamics in BP, as shown in Figure 1f. According to the fitting using bi-exponentially decaying function $\Delta T(t)/T = A_1 \exp(-t/\tau_1) + A_2 \exp(-t/\tau_2)$, the carrier relaxation time consists of two components, a fast decay time $\tau_1 = 26.26 \pm 1.8$ fs and a slow decay time $\tau_2 = 1.87 \pm 0.8$ ps, indicating the two dynamic steps of photo-excited carriers. It should be noted that the fast component is not precisely resolved due to the limitation of the pulse width (35 fs) of the pump-probe laser. According to previous study on monolayer TMDs which have similar direct band gap[21, 31], the fast component is attributed to defect-assisted cooling or carrier-carrier scattering, and the slow component is related to the carrier-phonon scattering and interband recombination. We suggest that the two components of decay time in BP might have similar origins, which is worthy further systematical investigations. It should be noted that the carrier dynamics in BP is much faster than that in either monolayer or bulk TMDs[21]. The very fast relaxation processes verify the nature of the direct bandgap in our BP samples, which underpin the following pulsed laser applications.

**2.2 BP-polymer composites**

We incorporated BP nanoflakes into polymer matrices which can not only protect BP from water-induced layer-by-layer etching[32] but also increase the processability and compatibility to photonic components. Here, we proposed two approaches to fabricate BP-polymer composites which can be easily transferred onto optical mirrors or optical fiber facets as photonic devices. In the first approach, a sandwiched structure in which BP was placed in between two pieces of Poly(methyl methacrylate) (PMMA) thin films was prepared, as shown in Figure 2a. The resulting PMMA-BP-PMMA composite film was peeled off by the scotch tape and attached onto the end facet of optical fiber ferrule for further optical studies (see Figure S1, S2 in supporting information). This method is simple but lack of control and repeatability.

In order to make a composite in which BP nanoflakes are more uniformly distributed in polymer matrix, a semi-industry approach of electrospinning was used to prepare robust networked membranes consisting of electrospun BP/PVP nanofibers (Figure 2b). A transparent thin film can be produced with good uniformity by electrospinning (see inset of Figure 2f). The morphology of electrospun thin film is characterized by scanning electron microscope (SEM) (Figure 2c) and high density of nanofibers without beading is observed. High magnification SEM image in Figure 2d reveals that the nanofibers have smooth surface with uniform diameter around 200 nm. The dark contrast in the nanofiber (see TEM image in Figure 2e) refers to BP nanoflakes, suggesting the successful encapsulation of BP in PVP. The absorption spectra of BP-PVP composite and a reference PVP film are presented in Figure 2f. It is found that the BP-PVP composite has enhanced absorption in the near-infrared wavelengths in comparison to the pure PVP. The absorption difference ($\Delta\alpha$) between BP-PVP and pure PVP (see Figure S3b in Supporting Information) reveals a broad absorption peak at around 1200 nm, suggesting that the enhanced absorption originates from the encapsulation of BP in the composite.

**2.3 Saturable absorption**

The nonlinear absorption of BP composites was studied by two-terminal power-dependent experiments (see Figure S4 in Supporting Information). The representative result from PMMA-BP-PMMA polymer composites is shown in Figure 3a. It is found that the transmission is increased with the increase of incident light intensity due to saturation of absorption. Similar to graphene-based saturable absorbers[4,5], the absorption bleaching is originated from Pauli blocking process in which a large amount of photogenerated carriers cause band filling, as illustrated by the inset of Figure 3a. The saturable absorption in BP can be fitted by

$$\alpha(I) = \frac{\alpha_s}{1 + I/I_s} + \alpha_{NS} \tag{1}$$

where $\alpha_S$ and $\alpha_{NS}$ are the saturable and nonsaturable absorption, $I_s$ is the saturation intensity, defined as the optical intensity required in a steady state to reduce the absorption to half of its unbleached value. A saturation intensity of 1.53 MW cm$^{-2}$ can be obtained, which is comparable to those reported for graphene and semiconductor saturable absorber mirrors (SESAMs)[33]. The modulation depth of BP is found to be around 10.6 %, comparable to that of carbon nanotube-based saturable absorbers which also has resonant absorption in telecommunication bands[34,35]. These results suggest that BP-polymer composite is a good candidate for noise suppressor and saturable absorber.

**2.4 Laser application**

Similar to the graphene-polymer composite[36], BP-PVP composite film can be easily transferred and attached onto the end-facet of optical fiber ferrule (inset of Figure 2f and Figure S3a). Following the transfer, BP-covered optical fiber was incorporated into the fiber laser cavity shown in Figure 3b. In a standard operation, we first excluded the possibility of self-Q-switching of the fiber laser. Without using the BP device, only continuous-wave (CW) emission was obtained even when we increased the pump power from the start oscillation threshold to the maximum and/or tuned the angle of polarization from 0 to 360 degree by adjusting the polarization controllers (PCs). After inserting the BP saturable absorber device, Q-switching state occurred at an incident pump power of only 25 mW, which is a very low threshold value in comparison to those of graphene[37,38] and TMDs-based saturable absorbers[39,40].

The results of pulse generation with BP-PVP saturable absorber at the pump power of 140 mW are shown in Figure 4a-d. Typical Q-switching output spectrum is shown in Figure 4a. It can be seen that the fiber laser operated with the central wavelength of 1561.9 nm. Its 3 dB bandwidth is 1.5 nm. Figure 4b shows the typical pulse train with uniform intensity distribution which reveals a repetition rate of 23.48 kHz, corresponding to a time interval of 42.5 μs. The intensity distribution is very uniform without modulation, indicating good noise suppressing capability of BP-PVP composite. Figure 4c is the corresponding single pulse profile in time domain with a narrower sweep span. The pulse has a full width at half maximum (FWHM) of 4.35 µs with symmetric intensity profile. We have measured the RF spectrum (Figure 4d) and found that the signal-to-noise ratio (SNR) is over 53 dB, indicating high quality of the output pulse. Moreover, apart from the fundamental and harmonic frequency, we did not observe other frequency component in the RF spectrum with wider span (see inset of Figure 4d), confirming high stability of the Q-switched pulse generation. The Q-switching state induced by BP saturable absorber was further verified by the stable Q-switching pulse output at different pump power from 70 mW to 280 mW (see Figure S5).

Unlike mode-locked state in which the repetition rate and pulse duration is fixed by the cavity length, Q-switched lasers have power-dependent repetition rate and pulse duration. Figure 4e reveals the relationship between pulse duration as well as repetition rate and the pump power. With the increase of the pump power from 28 mW to 280 mW, the repetition rate increases linearly from 7.86 kHz to 34.32 kHz. The pulse duration significantly decreases when the pump power increases in the lower power range and becomes nearly constant when the pump power is higher than 50 mW. The shortest pulse width we observed is 2.96 µs. The output power as a function of pump power is shown in Figure 4f (black trace). The slope efficiency, *i.e.*, the slope of the line obtained by plotting the laser output power against the input pump power, is ~2%. The step-like fluctuation may result from the deformation of the composite thin film due to heating effect. The highest output power is limited by the maximum input pump power. The continuous operation without BP as saturable absorber is also shown in Figure 4f (blue trace). The output power under Q-switching state is much smaller than that under continuous wave state due to the absorption or scattering of BP composite. Considering the corresponding repetition rate, single pulse energy reaches as high as ~194 nJ with 6.67 mW output power. The pulse energy can be even higher with the higher pump power. In order to evaluate the long term stability of BP-polymer composites, the output optical spectra were recorded every hour for 5 hours at a fixed pump power of 200 mW (Figure S6). Neither the central wavelength drift nor new wavelength component was observed during the measurements, indicating excellent repeatability and stability. Besides Q-switching results, we also observed mode-locking state using drop-casted BP as saturable absorber (see Figure S7 in Supporting Information). However, limited by the high loss of BP aggregates, the mode-locking threshold is very high (beyond 100 mW), and the heating-effect-induced instability becomes prominent.

## 3. Conclusion

In conclusion, we have demonstrated BP-polymer composite as saturable absorber for pulse generation in fiber laser. BP shows very fast carrier dynamics and BP-polymer composite has a modulation depth of 10.6%. A highly stable Q-switched pulse generation was achieved and the single pulse energy of ~194 nJ was achieved. The incorporation of BP nanoflakes into polymer matrix not only protects BP from ambient environments but also produces a new type of optical material by combining the special optical properties of BP with the structural properties of the polymer. By choosing suitable polymers and optimizing the fabrication technique, many more functional composite materials based on BP can be developed for a wide range of photonic and optoelectronic applications.

## 4. Experimental Section

*BP synthesis:* BP crystals were purchased from Smart Elements. The solution-based mechanical exfoliation process started from the mixing of BP crystals with ethanol solution, followed by sonication in an ultrasonic bath (400 W) at room temperature for 43 hrs. The resulted BP solution was purified by centrifugation, with a rate of 4000 rpm for 60 min, to remove larger particles. The final purified BP-ethanol mixture has a concentration ~6 mg/mL.

*Fabrication of composites:* In order to make PMMA-BP-PMMA sandwiched structure, a thin layer of PMMA was formed on $SiO_2$ substrate by spin-coating with a rotational speed of 2000 r/min. Following purified BP solution was drop-casted onto the PMMA film and another layer of PMMA was further spin-coated (rotational speed of 2000 r/min) to cover BP after the evaporation of solvent. A scotch tape with a hole (3 mm in diameter) in the central was aligned with the BP sample area, and used to peel off the PMMA-BP-PMMA composite film from $SiO_2$ substrate. The whole area with suspended BP-polymer film was then attached onto the fiber end facet to form the saturable absorber device (see Supporting Information for more details).

Polyvinylpyrrolidone (PVP) (molecular weight, $M_n$ = 1300 000) was purchased from Sigma-Aldrich. 1.5 g PVP was dispersed in 8 mL ethanol and 5mL N,N-Dimethylformamide (DMF). After stirring for 24 h at 60 °C in sealed bottles, the wet paste was mixed with 12 mL purified BP solution and sonicated for 15 min before electrospinning. The optimized electrospinning parameters include an applied voltage of 14 kV, feeding rate of 0.8 mL $h^{-1}$ and needle-tip-to-collector distance of 120 mm. During electrospinning, the needle was fixed above a collector made of aluminum foils for 15 min to form a membrane with a diameter around 150 mm. Similar scotch-tape-assisted peeling off and transferring procedures were used to make BP-PVP saturable absorber device.

*Characterization:* Raman spectra were collected on a micro-Raman system (Horiba Jobin Yvon HR800) equipped with a 488 nm laser. The spot size of excitation laser is ~1 μm. The thickness and morphology of BP nanoflakes were studied by AFM (Veeco Dimension-Icon). The microstructure of BP was studied on a scanning TEM (JEOL JEM-2100F, operated at 200 kV) and the TEM samples were prepared by drop-casting BP dispersion onto carbon grids. The microstructure of BP-PVP composite was investigated on SEM (HITACHI S-4800) and TEM (JEOL JEM-2010FEF). The TEM samples were prepared by directly electrospinning the BP-PVP nanofibers on TEM grids.

*Pump-probe measurement setup:* The transient dynamical response measurements were performed using a femtosecond pulse laser with pulse duration of 35 fs and repetition rate of 2 kHz. Both the pump light and the probe light are at 1550 nm. The pump incident light power is $3.3 \times 10^3$ GW cm$^{-2}$ and the probe incident light power is $0.5 \times 10^3$ GW cm$^{-2}$. The laser beam was focused by a lens with the focus distance of 250 mm. Any coherent artifact on the transient signal was eliminated by using the cross-polarized configuration. It has been examined that the probe beam alone could not cause any nonlinear effect in our experiments.

*Fiber laser cavity:* A piece of 3 m Erbium-doped fiber (LIKKI Er-16/125) as gain medium was reversely pumped by a 975 nm laser diode (975 LD) through a 980 nm/1550 nm wavelength division multiplexer (WDM). A polarization-independent isolator (ISO) was used to ensure the direction of light propagation. The cavity polarization state and intra-cavity birefringence was adjusted by polarization controller (PC). A 10 dB port of a coupler was employed as 10% output. The total cavity length is about 11.5 m. The output frequency-domain characteristics and the time-domain profile characteristics were simultaneously monitored by optical spectrum analyzer (Yokogawa AQ6370C), 500 MHz oscilloscope (Tektronix MDO3054) integrated with the RF spectrum analysis function (bandwidth: 3 GHz) and optical power meter (EXFO PM-1623).

**Acknowledgements**

We acknowledge the support from the National High Technology Research and Development Program of China (863 Program) (Grant No. 2013AA031903), the youth 973 program (2015CB932700), the National Natural Science Foundation of China (Grant No. 51222208, 51290273, 91433107), the Doctoral Fund of Ministry of Education of China (Grant No. 20123201120026). S. Lin acknowledges the support from the Postdoctoral Science Foundation of China (No. 7131701013), Hong Kong Scholars Program Funding (No. 7131708014&No. G-YZ36) and the postdoctoral early development program of Soochow University (No. 32317156 & No. 32317267).

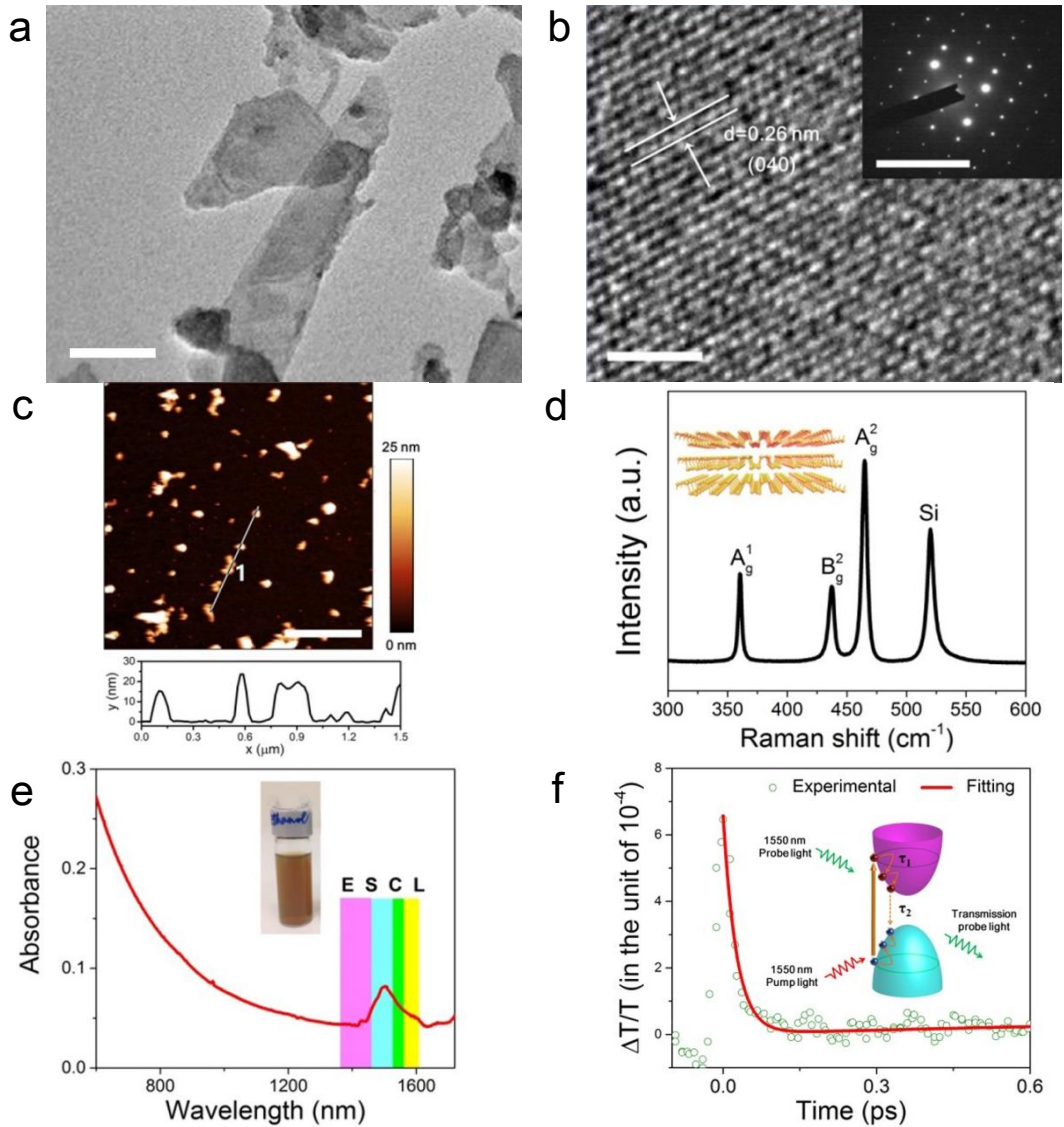

**Figure 1. Material characterizations of BP.** a) TEM image of BP nanoflakes. Scale bar: 50 nm. b) HRTEM image and of BP nanocrystal. Inset: SAED pattern. Scale bars are 1 nm and 10 nm$^{-1}$ (inset), respectively. c) AFM image of BP nanoflakes on SiO$_2$ substrate. Scale bar: 1 μm. The bottom figure shows the height profile along the white line. d) Raman spectrum of BP nanoflakes. The inset shows the atomic model of BP. e) UV-visible-infrared absorption spectrum of BP solution. The color bars indicate the telecommunication bands, *i.e.*, E: extended communication band; S: short wavelength communication band; C: conventional communication band; L: long wavelength communication band. The inset shows the photograph of the undiluted BP dispersion in ethanol. f) Measured transmittivity transients for BP solution. The open circles are the experimental data and the solid curve is the analytical fit to the data using exponentials with time constants $\tau_1$ and $\tau_2$. Inset schematically illustrates the transient dynamics processes in BP.

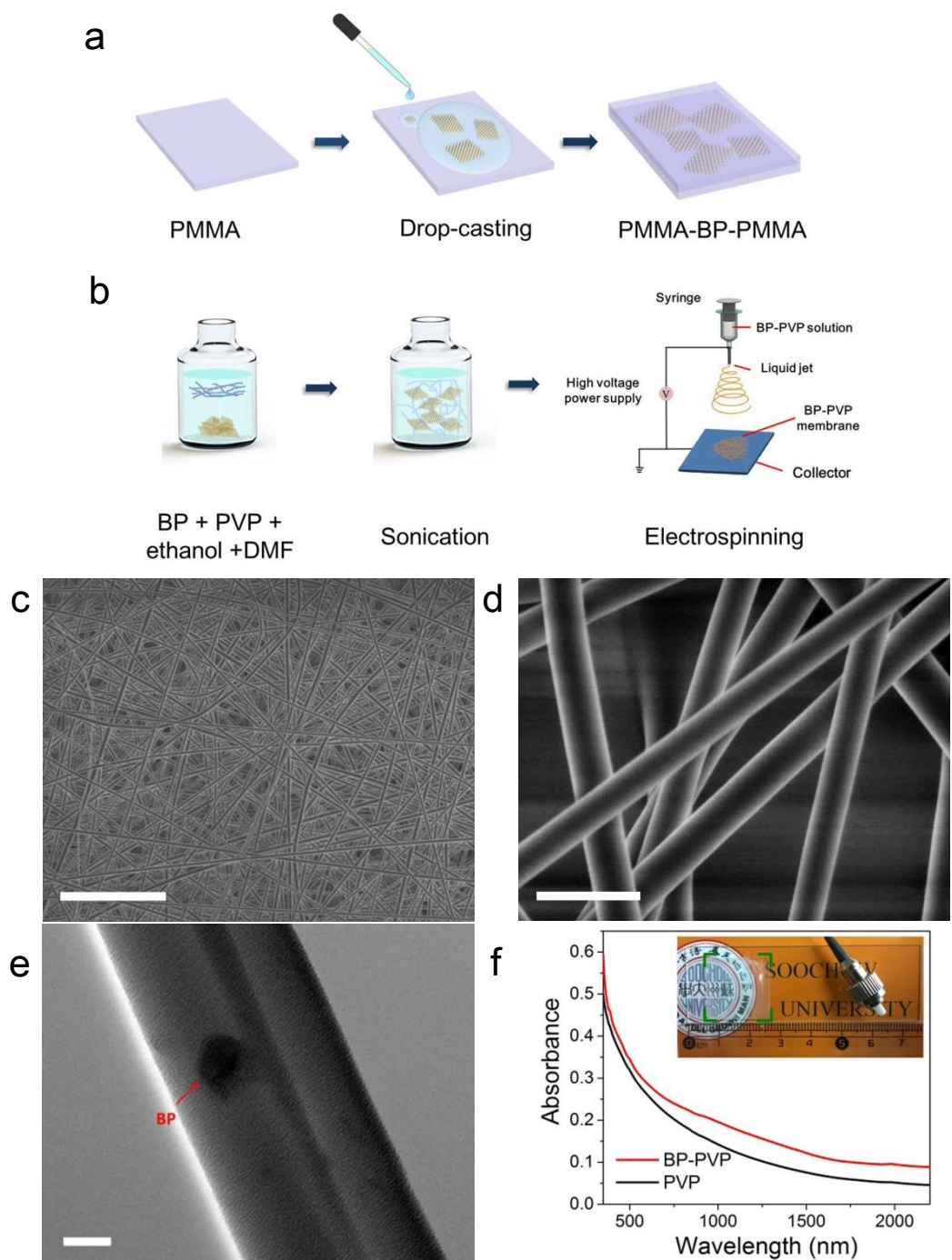

**Figure 2.** a) Schematics showing the fabrication process of sandwiched PMMA-BP-PMMA membrane. b) Schematics showing the fabrication of BP-PVP nanocomposite membrane by electrospinning. c) and d) SEM image of BP-PVP nanocomposite membrane fabricated by electrospinning. The scale bars are 5 μm (c) and 500 nm (d), respectively. e) TEM image of BP-PVP nanofiber composite. Scale bar: 50 nm. The red arrow indicates the BP nanoflake. f) UV-visible-infrared absorption spectrum of electrospun BP-PVP composite membrane (red line) and PVP membrane (black line). The inset shows the photograph of electrospun BP-PVP composite membrane on a quartz substrate (20 mm × 20 mm, left) as well as the optical fiber ferrule (right).

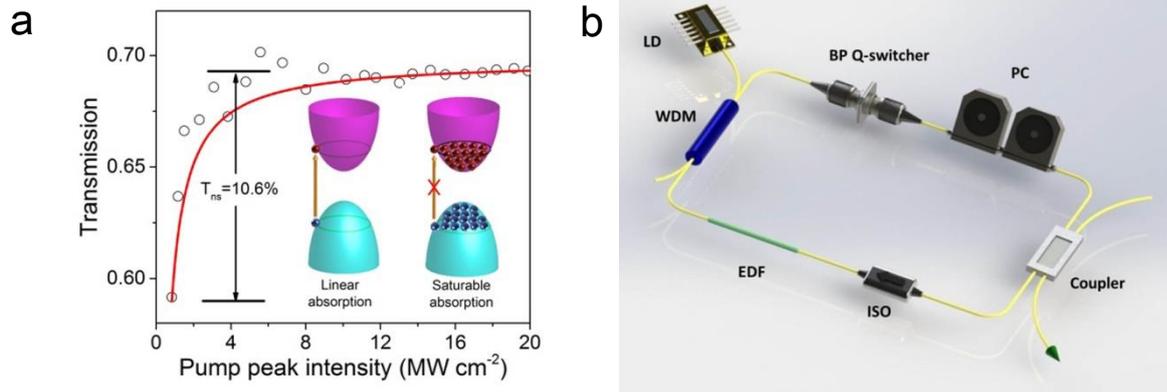

**Figure 3.** a) The saturable absorption curve of BP-polymer composites measured at 1565 nm. The insets show the energy diagrams of the linear absorption and saturable absorption. b) Schematic illustration of the ring cavity of the Q-switched fiber laser. LD: 974 nm laser diode; ISO: 1550 nm Polarization independent isolator; WDM: wavelength division multiplexer; PC: polarization controller; EDF: erbium-doped fiber.

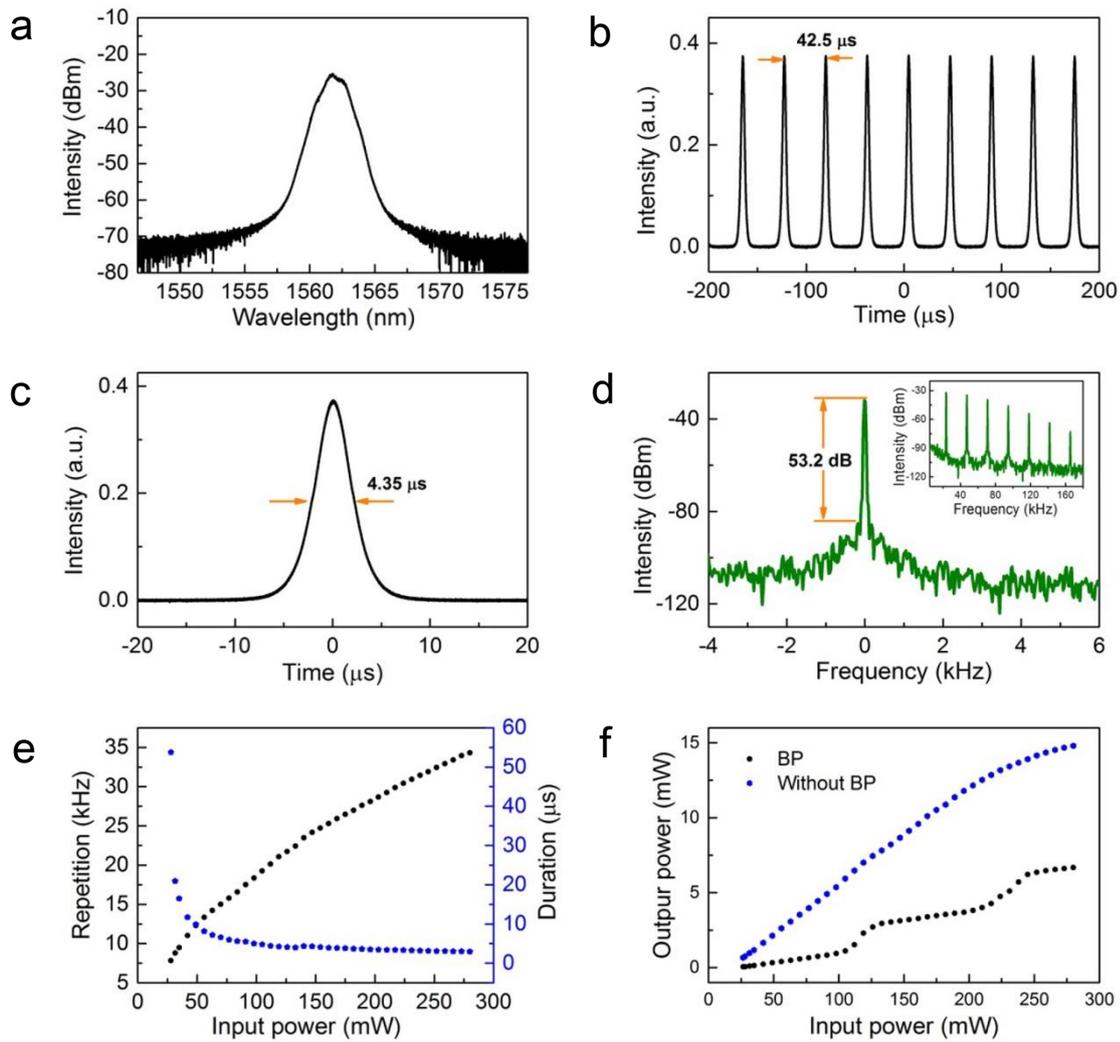

**Figure 4. Q-switching pulse output characterizations.** a) Optical spectrum. b) Q-switching pulse train. c) Single Q-switching pulse. d) The radio-frequency optical spectrum at the fundamental frequency and the wideband RF spectrum (inset). e) Pulse repetition rate and duration versus incident pump power. f) Output power versus incident pump power.

# Supporting Information

**Black Phosphorus-Polymer Composites for Pulsed Lasers**

*Haoran Mu, Shenghuang Lin, Zhongchi Wang, Si Xiao, Pengfei Li, Yao Chen, Han Zhang, Haifeng Bao, Shu Ping Lau, Chunxu Pan, Dianyuan Fan, Qiaoliang Bao\**

H. Mu, Dr. S. Lin, P. Li, Y. Chen, Prof. Q. Bao
Institute of Functional Nano and Soft Materials (FUNSOM), Jiangsu Key Laboratory for Carbon-Based Functional Materials and Devices, and Collaborative Innovation Center of Suzhou Nano Science and Technology, Soochow University, Suzhou 215123, P. R. China
(E-mail: qlbao@suda.edu.cn)

Dr. S. Lin, Prof. S. P. Lau
Department of Applied Physics, The Hong Kong Polytechnic University, Hung Hom, Hong Kong SAR, China

Z. Wang, Prof. C. Pan
School of Physical and Technology, Wuhan University, Wuhan 430072, P. R. China

Prof. S. Xiao
Institute of Super-Microstructure and Ultrafast Process in Advanced Materials, School of Physics and Electronics, Hunan Key Laboratory for Super-Microstructure and Ultrafast Process, Central South University, Changsha 410083, China

Prof. H. Zhang, Prof. D. Fan
SZU-NUS Collaborative Innovation Centre for Optoelectronic Science & Technology, and Key Laboratory of Optoelectronic Devices and Systems of Ministry of Education and Guangdong Province, Shenzhen University, Shenzhen, China

Prof. H. Bao
School of Materials Science and Engineering, Wuhan Textile University, Wuhan 430200, China

**\***H. Mu, S. Lin and Z. Wang contributed equally to this work.

\**Correspondence to*: Prof. Q. Bao (E-mail: qlbao@suda.edu.cn)

**Keywords**: black phosphorus, pulse laser, polymer composite, saturable absorber, electrospinning

# 1. The transferring process of PMMA-BP-PMMA from SiO$_2$ substrate onto the end facet of optical fiber and optical image characterization

Figure S1 shows the transfer process of PMMA-BP-PMMA composite film from SiO$_2$ substrate onto the end facet of optical fiber ferrule. After the fabrication of the sandwiched structure membrane (schematically shown in Figure S1a), we have attached a piece of scotch tape (Figure S1b) onto the composite film. The scotch tape has a hole with a diameter of 3 mm in the center which is aligned to the region with BP. Then the scotch-tape-attached SiO$_2$ substrate is immersed into DI-water (Figure S1c). Water is introduced at one edge of the exposed hydrophilic SiO$_2$/Si that penetrates between the hydrophilic SiO$_2$/Si substrate and hydrophobic carrier polymer attached with nanomaterials and immediately separates them within several seconds.[1] As a result, we can easily peel off the PMMA-BP-PMMA composite film and transferred it onto fiber facet (Figure S1d). With the help of optical microscope (Leica DM 4000), we can easily align the BP sample with optical fiber core area, as shown in Figure S2. We can see that the 9 μm fiber core (in the center of 125 μm fiber cladding) is fully covered by BP aggregates.

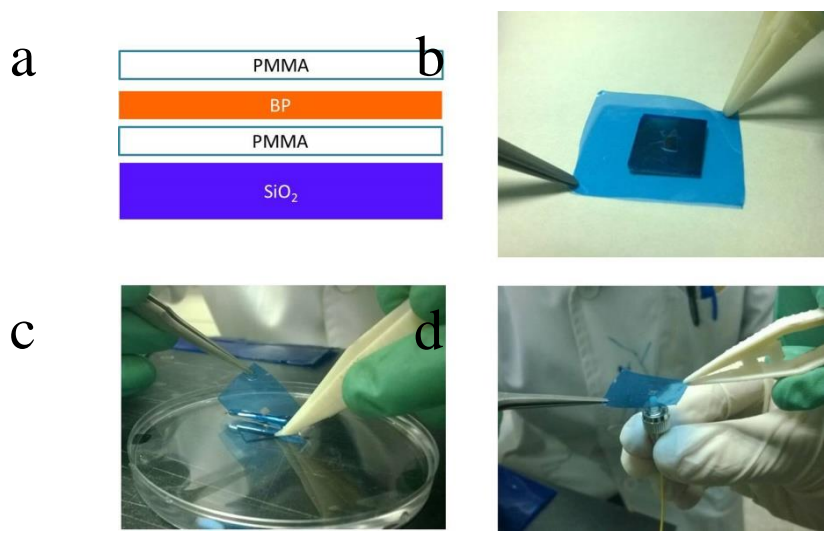

**Figure S1.** Photographs showing the transfer process of PMMA-BP-PMMA composite film onto the end facet of optical fiber. a) Schematic of the sandwiched composite thin film. b) A piece of scotch tape with a hole attached onto SiO$_2$/Si substrate. c) Peeling off the PMMA-BP-PMMA composite film from SiO$_2$ substrate in water. d) Transferring the composite thin film onto the end facet of optical fiber.

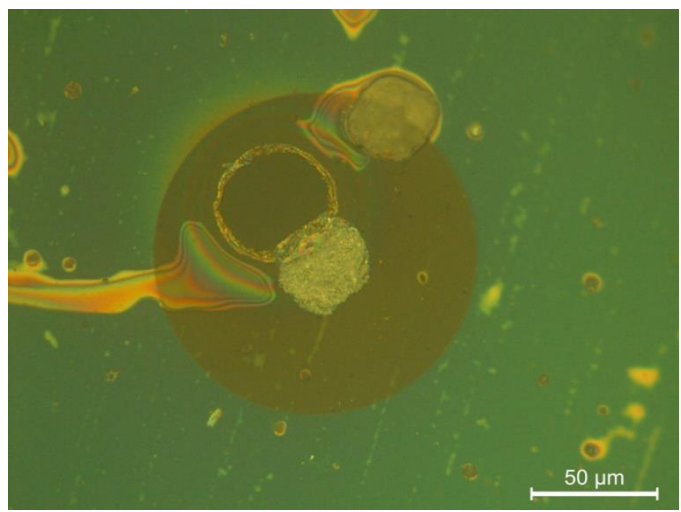

**Figure S2.** Optical image of PMMA-BP-PMMA composite film on the end facet of optical fiber ferrule. The fiber core area (9 μm area at the central) is fully covered by BP particles.

## 2. The characterization of electrospun BP-PVP composite membrane

The electrospun BP-PVP composite membrane can be peeled off and attached onto the end facet of optical fiber. Figure S3a shows the typical optical image of the end facet of optical fiber which is covered with the electrospun BP-PVP composite membrane. We can see that the fiber core area is fully covered with nanofiber network. Some liquid drop is also observed.

The absorption spectra of the BP-PVP composite and a reference PVP film are presented in Figure S3b. The absorption difference (Δα) between BP-PVP and pure PVP is presented in the inset of Figure S3b. The absorption peak of BP in composite is broadened and blue-shifted compared with that of BP in solution (Figure 1e). This is because the BP nanoflakes for electrospinning have smaller size and thickness, which were separated by centrifuging.

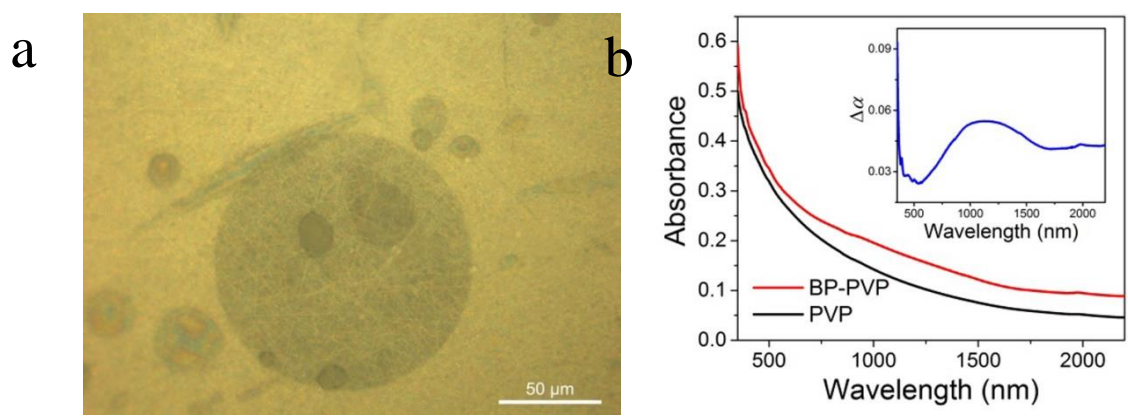

**Figure S3.** a) Optical image of BP-PVP matrix film on the fiber facet. The central area of the fiber core is completely covered by the composite film. b) UV-visible -infrared absorption spectrum of electrospun BP-PVP composite membrane (red line) and PVP membrane (black

line). The inset shows the different of the optical absorption curve between BP-PVP composite membrane and PVP membrane (Δα).

## 3. Nonlinear absorption measurements

In order to characterize the nonlinear optical response of the as-fabricated saturable absorber device, we have performed the balanced twin-detector measurements at telecommunication wavelength using the experimental setup as shown in Figure S4. A stable pulse train was emitted from a home-made passively mode-locked laser source. The laser has a repetition rate of 20 MHz with central wavelength at 1565 nm and 3 dB bandwidth of 6.9 nm. The pulse width is 407 fs, and the maximum power is 420 μW, corresponding to the maximum incident intensity of 20.622 MW cm$^{-2}$. An optical analyzer is used simultaneously to monitor the mode-locking state by a 5% coupler. We could directly control the input power by adjusting the tunable attenuator connected to the source. After passing through a 10% coupler, the output light is separated into two laser beams with almost identical strength. The first laser beam passed through the sample is measured by the photodiode detector A, whereas the second laser beam as a reference beam is directly measured by another photodiode detector B. These two detectors are connected with a dual-channel power meter (EXFO PM-1623), which synchronously measured the input and output pulse intensity to effectively reduce the detection errors. By continuously adjusting the incident power, the results of output power versus the input power are recorded by the power meter and consequently the optical transmittance under different input power can be calculated.

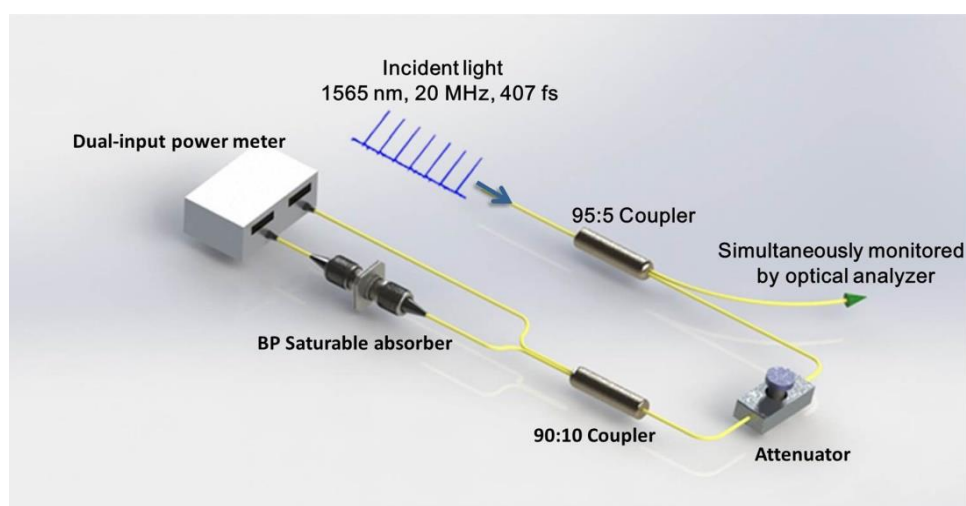

**Figure S4.** Schematic diagram showing the experimental setup for the nonlinear absorption measurements.

## 4. Q-switching state characterization under various pump power.

At a fixed cavity polarization, by continuously increasing the pump power, the pulse repetition is increased while the pulse train still remained as a uniform intensity distribution without obvious fluctuation, as shown in Figure S5. These experimental results indicate good stability of the passively Q-switched fiber laser enabled by the BP-based saturable absorber.

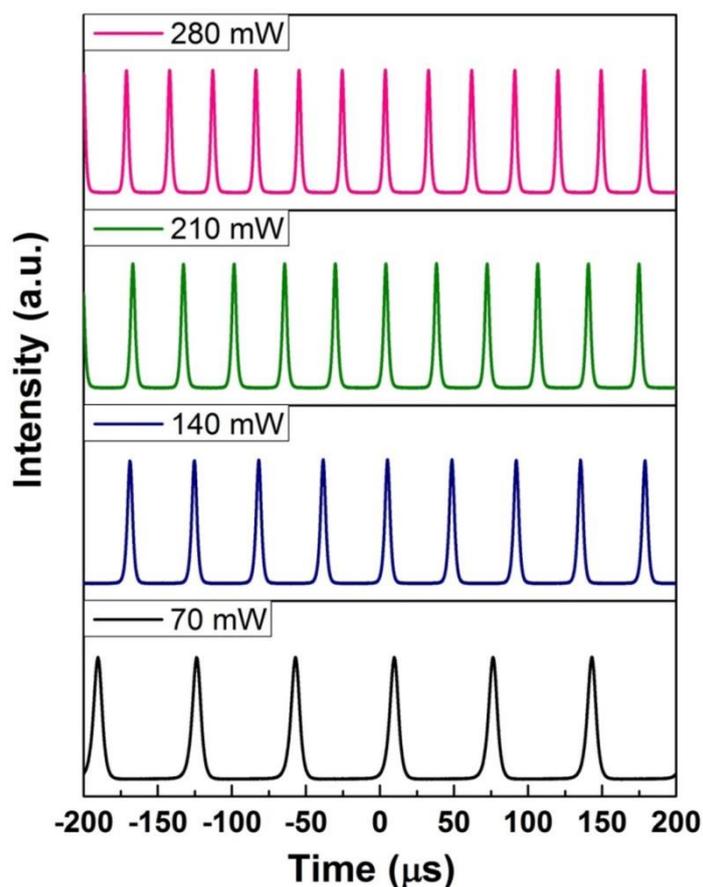

**Figure S5.** Pulse trains obtained at different pump powers.

## 5. Long-term stability of BP-based Q-switched fiber laser

To investigate the long-term stability of the Q-switched pulse laser, we recorded the optical spectra every hour over 5 hours at a fixed pump power of 200 mW, as shown in Figure S6. During our measurements, the center-wavelength is stable at 1563.5 nm and the 3dB bandwidth is around 1.5 nm. Neither noise wave nor continues wave was observed during our measurements, suggesting that the Q-switched laser possesses a reasonably good stability that is suitable for practical applications.

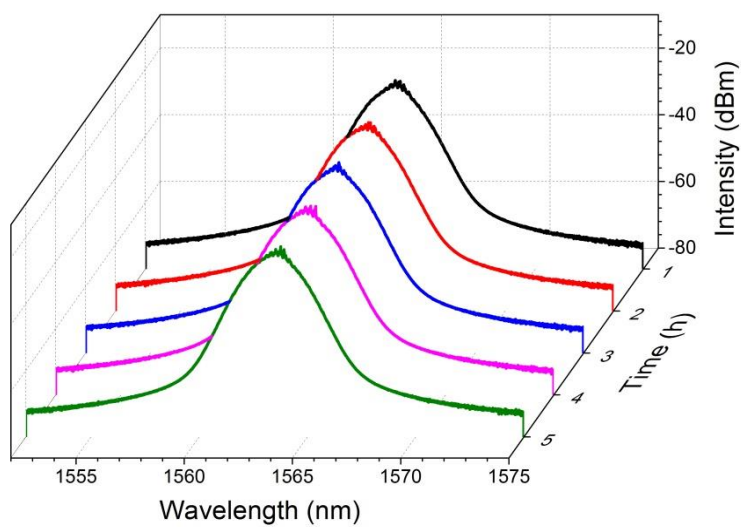

Figure S6. Output optical spectra of the Q-switched fiber laser as a function of operation time.

**6. Mode-locked pulse generated from BP-based saturable absorber**

Mode-locking state can be observed using drop-casted BP as saturable absorber. However, limited by the high loss of BP aggregates, the mode-locking threshold is very high (beyond 100 mW). Due to the degradation of BP in air, the heating-effect-induced instability becomes prominent. The mode-locking results are shown in Figure S7. The output spectrum has a symmetric and smooth shape without modulation wave, as shown in Figure S7a. The 3 dB bandwidth is 1.4 nm. However, no side-bands are observed, which indicates that the mode-locking pulse is not in soliton state. The non-soliton mode-locking is normal and has been widely reported in previous papers.[2-4] The pulse train in Figure S7b shows that the repetition rate of pulse train is 17.88 MHz. Due to the limit of the bandwidth of our oscilloscope, the pulse duration was measured.

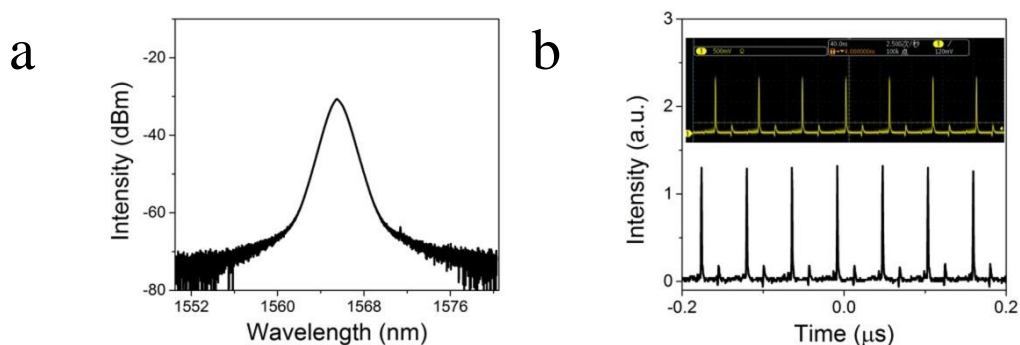

Figure S7. Mode-locked pulse generation from BP. a) Optical spectrum. b) Pulse train.